\documentclass[aps,prd,groupedaddress,]{revtex4}
\usepackage{latexsym}
\usepackage[dvips]{graphicx}

\newcommand{\eq}{\begin{equation}}
\newcommand{\feq}{\end{equation}}
\newcommand{\eqn}{\begin{eqnarray}}
\newcommand{\feqn}{\end{eqnarray}}
\newcommand{\arr}{\begin{eqnarray*}}
\newcommand{\farr}{\end{eqnarray*}}
\newcommand{\beq}{\begin{equation}}
\newcommand{\eeq}{\end{equation}}
\newcommand{\bea}{\begin{eqnarray}}
\newcommand{\eea}{\end{eqnarray}}

\def\beq{\begin{equation}}
\def\eeq{\end{equation}}
\def\bea{\begin{eqnarray}}
\def\eea{\end{eqnarray}}
\def\bc{\begin{displaymath}}
\def\ec{\end{displaymath}}
\def\lb{\label}

\def\be{\beta}

\def\la{\lambda}
\def\laq{\lambda^{2}}

\def\lb{\label}

\begin{document}

\preprint{INFNCA-TH0304}

\title{Black Hole fragmentation and  holography}

\author{M.~Cadoni}
\email{mariano.cadoni@ca.infn.it}

\affiliation{Dipartimento di Fisica,
Universit\`a di Cagliari, and INFN sezione di Cagliari, Cittadella
Universitaria 09042 Monserrato, ITALY}

\begin{abstract}

We discuss the  entropy change due to  fragmentation for black hole 
solutions in various dimensions. We find three different types of behavior. 
The entropy may decrease,  increase or have a mixed 
behavior, characterized by  the presence of a threshold mass. 
For two-dimensional (2D) black holes we give  a complete 
characterization of the entropy behavior under fragmentation, in the form of 
sufficient 
conditions imposed on the function $J$, which defines the 2D 
gravitational model.  We compare the behavior of the gravitational 
solutions with that of free field theories in $d$ dimensions. This 
excludes the possibility of finding a gravity/field theory 
realization of the holographic principle for a broad class of 
solutions,
including asymptotically flat black 
holes. We find that the most natural candidates for holographic duals 
of the black hole solutions with mixed behavior are field theories 
with a mass gap. We also discuss the possibility of formulating 
entropy bounds that make reference only to the energy of a system.

\end{abstract}


\maketitle

\section{Introduction}

One of the most striking novelties in the research on gravitational 
physics is the possibility that gravity in $d$ dimensions could be 
described by a local field theory in $d-1$ dimensions 
\cite{Maldacena:1997re,Witten:1998qj,Gubser:1998bc}. The theoretical
evidence 
for such holographic description of gravity is mounting. Indications 
that holography could be a fundamental feature of the gravitational 
interaction come from different directions: string theory, black
hole physics, cosmology \cite{'tHooft:gx,Susskind:1994vu,Verlinde:2000wg}
(For a recent review see \cite{Bousso:2002ju}).  
A particularly interesting output of these 
investigations has been the formulation of stringent holographic bounds 
for the  entropy  of  a system occupying a given region of 
space \cite{Susskind:1994vu,Bousso:1999xy}.

An explicit realization of the holographic principle has been 
found only  in particular cases, essentially for anti- de Sitter (AdS) 
(and de Sitter) gravity, the so-called anti de Sitter/conformal field 
theory (AdS/CFT) correspondence \cite{Maldacena:1997re,
Witten:1998qj,Gubser:1998bc}. 
A general way to explicit realize the holographic principle for generic 
gravitational systems, in particular for gravity in asymptotically 
flat spacetimes,  is still lacking. In particular, it is not clear if 
the realizations of the  holographic principle take always the 
form a correspondence between $d$-dimensional gravity and a field theory 
in $d-1$ 
dimensions, or if there could be some alternative, still unknown, 
realization of it.
If  the holographic principle has to be considered   a genuine feature 
of every  quantum theory of gravity, one could explain our lacking in  
understanding  the holographic principle as a lacking in 
understanding  quantum gravity. However, there are  strong indications 
that holography is a feature of gravity that appears and therefore 
should be explained,  already at the  semiclassical level. 
The Bekenstein-Hawking area law for the black hole entropy is the 
most striking example of holographic behavior of a gravitational 
system that has to be explained already at the semiclassical level.

An alternative strategy one can use in this context, is to explore 
the similarities and the differences between gravity and local field 
theories in order to check at a fundamental level the possibility of 
finding correspondences  between the two classes of theories.
This approach can be very powerful. One nice example, discussed in 
almost every introducing paper on the holographic principle, is the 
scaling behavior of the entropy  as a function of the volume of the 
system
for a local field theory compared to 
that of a black hole.  For a local field theory the entropy is an 
extensive quantity, it scales as the volume of the space. On the other 
hand the entropy of a black hole scales as the area of the horizon. 
This simple fact enables one to conclude 
that the correspondence between gravity   and field theory, if it 
exists, must 
be {\sl{holographic}}.

In this paper we will focus on an other aspect of the relationship 
between gravity and local field theory, namely on the dependence for 
composite systems  of the entropy from the energy.  Working in the microcanonical 
ensemble  the entropy /energy relation for  a free field  theory in $d$-dimensions 
is given by
\beq\lb{enen}
S\propto E^{(d-1)/d},
\feq
which considering two arbitrary excitations
with energy $E_{1}, E_{2}$
satisfies the inequality 
\beq\lb{ineq}
S(E_{1}+E_{2})< S(E_{1})+S(E_{2}),
\feq
Conversely asymptotically flat black holes of masses $M_{1}, M_{2}$ in $d\ge 4$ dimensions 
satisfy
\beq\lb{ineq1}
S(M_{1}+M_{2}) > S(M_{1})+S(M_{2}),
\feq
i.e fragmentation of a black hole of mass $M_{1}+M_{2}$  into two 
smaller black holes of masses $M_{1}$ and $M_{2}$ is entropically 
not preferred. Assuming the existence of a gravity/field 
theory correspondence,  one has to identify the black hole masses $M_{1}$, $M_{2}$ as 
excitations $E_{1}$, $E_{2}$ of the field theory. It follows that Eq. 
(\ref{ineq}) contradicts Eq. (\ref{ineq1}) and consequently that the 
assumed correspondence gravity/field 
theory cannot be true.  Also skipping this problem, 
i.e assuming the existence of black hole solutions satisfying   
inequality (\ref{ineq}) rather then  (\ref{ineq1}),  one is faced by 
an other problem. At first sight Eq. (\ref{ineq}) seems incompatible 
with every entropy bound because a black hole could always increase 
its entropy by fragmentating into smaller black holes.

In this paper we will analyze in detail for various classes of black 
holes in various dimensions the validity of the relation (\ref{ineq1}).
We will show that Eq. (\ref{ineq1}) holds true only for asymptotically 
flat black holes. Black holes with different asymptotic  behavior 
(for instance AdS black holes) may satisfy Eq. (\ref{ineq}). 
For the two-dimensional (2D) case we will be able to give a complete 
characterization of the entropy for composite black holes in 
terms of the asymptotic behavior of 
the solution.  We will also show the existence of black hole solutions with 
mixed behavior, i.e solutions satisfying either Eq. (\ref{ineq}) or Eq. (\ref{ineq1}) 
when the  masses $M_{1}$ and $M_{2}$ are above or below  some critical 
value $\tilde M_{0}$.
We will argue that this mixed behavior as a natural counterpart in a 
field theory with a mass gap. Finally, we also discuss the 
compatibility of Eq. (\ref{ineq}) with entropy bounds.

\section{Asymptotically flat black holes}
Let us first consider the $d$-dimensional Schwarzschild black hole 
($d\ge 4$)
\beq\lb{schwar}
ds^{2}=- \left(1- {k_{d}¥GM\over r^{d-3}}\right)dt^{2}+ \left(1- {k_{d}GM\over 
r^{d-3}}\right)^{-1}dr^{2}+ r^{2}d\Omega^{2}_{d-2},
\feq
where $M$ is mass and $k_{d}=16\pi/(d-2)\Omega_{d-2}$,
$\Omega_{d-2} $ being the volume of the unit $S^{d-2}$ transverse sphere. 
The black hole entropy is given by the area law
\beq\lb{entropy}
S={A\over 4G}= {\Omega_{d-2}\over 4}(k_{d} L_{p}M)^{(d-2)/(d-3)},
\feq
where  $L_{p}=G^{1/(d-2)}$ is the Planck length.

In order to discuss the inequality (\ref{ineq1}) on a physical 
ground, we will make here and throughout the all paper the following 
assumptions: 1) The theory admits, at least in some  approximation, 
multi black hole solutions and 2) 
The gravitational potential energy of the multi black hole 
configuration can be neglected with respect to the black hole masses.
Assumption 1) is necessary in order to give to a multi black hole 
configuration a precise meaning whereas assumption 2) assures us 
that the multi black hole configuration can be treated as a 
composite system with zero binding energy, whose mass and entropy is 
simply the sum of that of the  elementary constituents.

Using Eq. (\ref{entropy}) one easily find that Eq. (\ref{ineq1}) is 
satisfied. As expected, for Schwarzschild black holes the entropy is 
maximized by the single 
black hole configuration with mass $M_{1}+M_{2}$.
The same holds  true for asymptotically flat  charged black holes. Considering for 
instance the four-dimensional Reissner-Nordstrom solution one finds for 
the entropy  
\beq\lb{entropy1}
S=\left(ML_{p}+\sqrt {M^{2}L_{p}^{2}-Q^{2}}\right)^{2}, 
\feq
where $Q$ is the electric charge of the black hole.
Again, from Eq. (\ref{entropy1}) it follows $S(M_{1}+M_{2})> 
S(M_{1})+S(M_{2})$.
There is  one simple argument that can be used to argue
that  inequality (\ref{ineq1}) holds in general for asymptotically 
flat black holes at least when $ML_{p}$ is much bigger then the 
(eventually present) black hole charges.
The area law  gives $S\propto (r_{+}/L_{p})^{d-2}$, where  $r_{+}$ is the horizon 
radius. For asymptotically flat black holes and when $ML_{p}>>Q_{i}$, 
where $Q_{i}$ are the charges associated with the black hole, the 
gravitational potential is dominated by the Newtonian term so that we have $ 
r_{+}\propto L_{p}^{(d-2)/(d-3)}M^{1/(d-3)}¥$. It follows $S\propto (L_{p}M)^{(d-2)/(d-3)}¥$, 

\section{ AdS black holes}

The argument presented at the end of the previous section does not 
apply to non-asymptotically flat black holes. For 2D black holes a general 
discussion we will presented  in the 
next section.  Here we will discuss 
only the most interesting case, namely asymptotically AdS black holes.
The entropy-mass relation for  the $d$-dimensional Schwarzschild-anti de Sitter 
black hole ($d\ge 4$), 
\beq\lb{schwarads}
ds^{2}= -\left(1+\la^{2}r^{2}- {k_{d}GM\over r^{d-3}}\right)dt^{2}+ 
\left(1+\la^{2}r^{2}-{k_{d}GM\over 
r^{d-3}}\right)^{-1}dr^{2}+ r^{2}d\Omega^{2}_{d-2},
\feq
is rather complicated.  However a simple formula can be found for 
black holes with $ML_{p}^{d-2}\la^{d-3}>>1$. In this latter  case we have
\beq\lb{en1}
S\propto \left( {M\over L_{p}\la^{2}}\right)^{(d-2)/(d-1)}.
\feq
Differently from the asymptotically flat case, we see that now 
inequality (\ref{ineq}) is satisfied. For AdS black holes ( at  least 
for those with large enough 
mass) fragmentation is entropically preferred. 
This fact is perfectly consistent with the existence of a AdS/CFT 
correspondence  between $d$-dimensional AdS gravity and 
$d-1$-dimensional conformal field theory. The $d$-dimensional black hole entropy (\ref{en1}) 
reproduces correctly, after the $E=M$ identification, 
the entropy (\ref{enen}) for a free field theory 
in $d-1$-dimensions.  

Let us now discuss in some detail the three-dimensional (3D) case, 
the Banados-Teitelboim-Zanelli (BTZ) black hole \cite{Banados:wn}.
This case is very 
instructive not only because it is possible to solve exactly the 
inequality (\ref{ineq}) but also because it clarifies the role played by 
the black hole  ground state in our considerations.
The BTZ black hole solution  with zero angular momentum is 
\beq\lb{btz}
ds^{2}= -(\la^{2}r^{2}- 8GM)dt^{2}+(\la^{2}r^{2}- 
8GM)^{-1}¥dr^{2}+r^{2}d\phi^{2}¥¥,
\feq
where the black hole mass $M$ is defined with reference to the 
$M=0$ black hole ground state,  $ds^{2}= 
-\la^{2}r^{2}dt^{2}+\la^{-2}r^{-2}¥dr^{2}+r^{2}d\phi^{2}$.
The entropy is given by $S=2\pi r_{+}/4G= (\pi/\la) \sqrt{2M/G}$ so that 
Eq. (\ref{ineq}) is satisfied for every value of the mass $M$.
On the other hand, if the  black hole mass is defined with reference to 
the full, geodetically complete AdS   spacetime 
$ds^{2}= 
-(1+\la^{2}r^{2})dt^{2}+(1+\la^{2}r^{2})^{-1}dr^{2}+r^{2}d\phi^{2}$,
the mass spectrum looks rather different. The $M=0$ ground state is 
separated from the continuos part of the  spectrum with $M\ge 1/8G$ 
by a mass gap. For $M\ge 1/8G$ the entropy is  given by $S=(\pi/2\la G) \sqrt{8GM-1}$.
Taking for simplicity $M_{1}=M_{2}=M$ we see that Eq. (\ref{ineq}) is 
satisfied only for $M\ge  3/16 G$, whereas for $1/8G \le M\le  3/16 G$ 
we have
$S(M_{1}+M_{2})\ge S(M_{1})+S(M_{2})$. We will come back to this 
point in Sect. V , where will argue that this behavior is 
typical for a spectrum with a mass gap.

\section{Two-dimensional black holes}
In $d\ge3$ spacetime dimensions it is very difficult 
to formulate general criteria that enable us to decide if a 
black hole  satisfies either Eq. (\ref{ineq}) or Eq. (\ref{ineq1}).
These criteria can be found  for 2D black holes. 
The 2D case is interesting for several reasons. Two-dimensional 
gravity  supports a realization of the AdS/CFT correspondence 
\cite{Cadoni:1998sg}. In two spacetime dimensions we can formulate entropy 
bounds \cite{Mignemi:2003uz}.  
Moreover, 2D black holes can be used to describe black holes in higher dimensions. 
Two dimensional 
black holes arise as effective description of the near-horizon, near-extremal 
behavior of $d$-dimensional charged black hole and branes.
Every  $d$-dimensional spherically symmetric solution can be described,
after dimensional reduction, by a 2D dilaton gravity model.

The generic 2D dilaton gravity  model (For a recent review see \cite{Grumiller:2002nm}) 
can be completely 
characterized by a dilaton potential $V(\Phi)$, the action for the 
model being given by $A=(1/2)\int d^{2}x \sqrt{-g}(\Phi R+\laq 
V(\Phi))$. 
The general 2D black hole solution takes the form
\beq\lb{2dbh}
ds^{2}=-\left( J(\Phi)- {2 M\over \la}\right)dt^{2}+\left( J(\Phi)- {2 
M\over \la}\right)^{-1}¥dr^{2},\quad \Phi=\la r,
\feq
where $J(\Phi)=\int V(\Phi)$  and $M$ is the black hole mass.
To be sure that Eq. (\ref{2dbh}) describes a black hole we will 
take  $M\ge 0$, $\Phi>0$  and $J(\Phi)>0$ a strictly increasing 
function of $\Phi$, (we will therefore have $J( \infty)=\infty$).  
The temperature and entropy associated with the black hole are given by 
\beq\lb{ts}
S= 2\pi \Phi_{h},\quad\quad T= {\la\over 4 \pi} V(\Phi_{h}),
\feq
where $\Phi_{h}= J^{-1}(M)$ is the value of the dilaton at 
the black hole horizon. 

Using Eq. (\ref{2dbh}) one easily finds 
\beq\lb{en2}
S=2\pi J^{-1}(M).
\feq
One can now  find sufficient conditions to be imposed on the function 
$J(\Phi)$ such that Eq. (\ref{ineq}) is satisfied  $\forall M_{1},M_{2}$.
Let us first show that if both conditions (the prime denotes 
derivation with respect to $M$)
\bea\lb{conda}
&&(J^{-1})''(M)<0,\quad \forall M,\\
\lb{condb} &&J^{-1}(M=0)=0,
\eea
are satisfied then
\beq\lb{eq1}
J^{-1}(M_{1}+M_{2})<J^{-1}(M_{1}) +J^{-1}(M_{2}¥),
\feq
which, owing to Eq. (\ref{en2}) implies that the inequality (\ref{ineq}) 
for the entropy is also satisfied.
The convexity condition (\ref{conda})  implies for every positive $\hat 
M, M_{1},$ 
$(J^{-1})'(\hat M +M_{1})<(J^{-1})'(\hat M )$, from which it  follows 
$J^{-1}(\hat M +M_{1}+M_{2})+J^{-1}(\hat M) < J^{-1}(\hat M +M_{1})+J^{-1}
(\hat M +M_{2})$.
Evaluating the previous inequality at $\hat M=0$ and using condition 
(\ref{condb}) one easily recovers Eq. (\ref {eq1}). 

Analogously, we can show that 
if  conditions
\bea\lb{conda1}
&&(J^{-1})''(M)>0,\quad \forall M,\\
\lb{condb1} &&J^{-1}(M=0)=0,
\eea
hold then the inequality 
(\ref{ineq1}) is satisfied for every $M_{1},M_{2}$.
If $(J^{-1})''=0$, identically, then the black hole solution of the 
model will satisfy $S(M_{1}+M_{2})=S(M_{1})+S(M_{2})$.
If $(J^{-1})''$ changes sign, we cannot make any 
definite statement about inequalities (\ref {ineq}) or (\ref {ineq1}).

Any function $J^{-1}(M)$  which is everywhere convex (concave) and  
strictly  growing must diverge for 
$M\to\infty$ lesser (faster) then $J^{-1}= M$. It follows that $J(\Phi)$ must 
diverge for $\Phi\to\infty$   faster  (lesser) then $J= \Phi$.
We have reached an important result:
for models satisfying the required criteria about the derivatives of 
$J^{-1}$ and $J(0)=0$,   fragmentation of 2D black holes 
can be  or not be entropically preferred depending on the asymptotic 
behavior of the function $J$. 
The black solutions of  the 2D dilaton gravity model,
characterized by a function $J$, which asymptotically diverges  
faster (lesser) then $\Phi$, 
 will satisfy Eq. (\ref{ineq}) (Eq. (\ref{ineq1})).

When $J^{-1}¥(0)\neq 0$ but condition (\ref{conda}) still holds we will have a mixed 
behavior. The black hole will satisfies Eq. (\ref {ineq}) for $M> 
\tilde M_{0}$ and  Eq. (\ref{ineq1}) for $M< \tilde M_{0}$, where 
$\tilde M_{0}$ is some threshold mass.
Let us first notice that only the case
$J^{-1}¥(0)=\Phi_{0} <0$ has physical relevance. Being $J^{-1}$ 
a strictly growing function, $\Phi_{0}>0$ implies $J^{-1}¥( M_{0})=0$ with $ 
M_{0}$ {\sl negative}, i.e
the presence of negative 
masses in  the  spectrum.
Conversely, $\Phi_{0}<0$ implies 
\beq\lb{en3}
J^{-1}¥( M_{0})=0,
\feq
with $ 
M_{0}$ {\sl positive}, i.e
the black hole spectrum  is limited from below by the extremal, non 
vanishing  value of the mass  $M_{0}$. In general the extremal mass 
$M_{0}$ is 
simply related with the threshold mass $\tilde M_{0}$.
Using Eqs. (\ref{en2}) and (\ref{en3})  one easily realizes that the entropy of the 
extremal  state is zero, $S(M_{0})=0$. Depending on the behavior of 
the dilaton potential at $M_{0}$ we can have two cases.
1) $J^{-1}¥( M_{0})=0$ and $V( M_{0})=0$. The extremal state has zero 
entropy and from Eq. (\ref{ts}) also zero temperature.
2) $J^{-1}¥( M_{0})=0$ and $V( M_{0})\neq 0$.
The extremal state has zero 
entropy but nonvanishing temperature. In the next section we shall see
that in both cases  this behavior can be explained with the 
presence of a mass gap.

Let us now give some examples to illustrate our 
general results. The simplest model satisfying both conditions 
(\ref{conda}),(\ref{condb})  
 is the Jackiw-teitelboim model, $V=2\Phi$. We have 
$J^{-1}(M)=\sqrt{2M/\la}$,  which satisfies both $ (J^{-1})''<0$ and
$J^{-1}(0)=0$. For the entropy we get $S\propto \sqrt{M}$, from which 
Eq. (\ref{ineq}) follows for every value of the mass.
A more general model is given by \cite{Mignemi:1994wg}
\beq\lb{en4}
V= (h+1)\Phi^{h}, \quad h>-1.
\feq
For $h>0$ the model satisfies the condition (\ref{conda}) and (\ref{condb}). The 
black hole entropy is given  by 
\beq\lb{en7}
S\propto (M/\la)^{1/(h+1)},
\feq
which as 
expected satisfies Eq. (\ref{ineq}). This class of models contains, 
as particular cases, 2D gravity models arising as near-horizon limit 
of dilatonic zero-branes \cite{Cadoni:2001ew},    black 3-branes \cite{Cadoni:2003vi}
and heterotic string black holes \cite{Cadoni:1999gh}. 
It is interesting to note that the entropy-energy relation 
(\ref{en7}) becomes that  of a free field theory given by Eq. (\ref{enen}) 
identifying $h=1/(d-1)$. This fact gives an hint about the possibility 
of finding a correspondence between these 2D dilaton gravity models 
and a free field theory.

For $-1<h<0$ the model satisfies the conditions (\ref{conda1}) and 
(\ref{condb1}). The 
black hole entropy now satisfies Eq. (\ref{ineq1}). An important 
particular case is given  by $h=-1/2$, which describes the spherical 
dimensional reduction of the Schwarzschild black hole. 
For $h=0$ we have the CGHS model \cite{Callan:rs} . Eq. (\ref{en4}) 
gives $(J^{-1})''=0$ identically. The 
entropy  depends  linearly on  the mass, so that, as expected, 
$S(M_{1}+M_{2})=S(M_{1})+S(M_{2})$. 

As an examples of a model  fulfilling  condition (\ref{conda}) but not
(\ref{condb}) let 
us first consider the exponential potential $V=\be\exp(\be \Phi)$, 
with $\be>0$. The black hole horizon is located at $\la r=\Phi=J^{-1}(M)= 
(1/\be)\ln(2M/\la)$. $J^{-1}$ diverges  asymptotically lesser then 
$M$ but we have  $J^{-1}(0)\neq 0$. We therefore expect the 
presence of a threshold mass $\tilde M_{0}$ separating the two regions
of the spectrum where  
Eq. (\ref{ineq}), respectively Eq. (\ref{ineq1}) hold.

The black hole mass has an extremal 
value,  $M\ge M_{0}=\la/2$. However, 
$J^{-1}(M_{0})=0$ and $V(M_{0})\neq 0$. The entropy and temperature 
of the black hole are
\beq\lb{en5}
S= {2\pi\over \be}\ln{2M\over\la},\quad T = {\be\over 2\pi} M.
\feq
The ground state has zero entropy but finite temperature 
$T(M_{0})=(\be/4\pi)\la$.  Using Eq. (\ref{en5}) an taking for 
simplicity $M_{1}=M_{2}=M$, one can easily find that the inequality 
(\ref{ineq}) is satisfied  only for $M>\tilde M_{0}=\la$, whereas for 
$\la/2<M<\la$ Eq. (\ref{ineq1}) holds. Black hole fragmentation becomes 
entropically preferred for  $M$ bigger then the threshold mass $\tilde 
M_{0}$. Notice that the threshold mass $\tilde M_{0}$ although of the same 
order do not coincide with the extremal mass  $ M_{0}$,
we have $\tilde M_{0}= 2 M_{0}$. 

As a second example of models with $J^{-1}(0)\neq 0$ let us consider 
$J(\Phi)=\Phi^{2}+1$. This is the 2D analogue of the 3D BTZ black 
hole discussed at the end of the previous section. Also in this case the black hole 
mass has an extremal value $M_{0}=\la/2$. However, now $S\propto 
T\propto
\sqrt{(2M/\la)-1}$, so that the extremal state has zero entropy and 
temperature, $S(M_{0})=T(M_{0})=0$. The threshold mass at which black 
hole fragmentation becomes entropically preferred is given by $\tilde 
M_{0}= (3/2) M_{0}= (3/4)\la$.

Until now we have  considered a system in which the 
total energy is constant,  i.e fragmentation of a black hole of mass $M$ 
into two smaller black holes of masses $M_{1}+M_{2}=M$.
Let us now consider  systems  
at  constant volume. We  will show that  
the energy is minimized when the 
conditions which maximize the entropy are satisfied.

We have to consider the black hole mass $M$ as  a function of its radius 
$R$ and to solve the inequality
\beq\lb{ener}
M(R_{1}+R_{2})> M(R_{1})+M(R_{2}).
\feq
Using Eq. (\ref{2dbh}) one finds $M(R)=(\la/2) J(\Phi_{h})=(\la/2)
J(R)$. Eq. (\ref{ener}) becomes $J(R_{1}+R_{2})> J(R_{1})+J(R_{2})$.
Because $(J^{-1})''<0$ and $J^{-1}(0)=0$ imply, respectively, 
$J''>0$ and $J(0)=0$ it follows that  conditions (\ref{conda}), (\ref{condb}) 
for the function $(J^{-1})$ are equivalent to conditions 
(\ref{conda1}), (\ref{condb1})
for the function $J$. As a consequence, whenever conditions 
(\ref{conda}) and  (\ref{condb})
 are satisfied we have not only $S(M_{1}+M_{2})< S(M_{1})+S(M_{2})$
but also $M(R_{1}+R_{2})> M(R_{1})+M(R_{2})$. The process of fragmentation of a black 
hole maximizes the entropy if the total mass is constant and 
minimizes the mass if the total volume is constant.
Conversely, if conditions (\ref{conda1})  and  (\ref{condb1}) hold we have 
$S(M_{1}+M_{2})>S(M_{1})+S(M_{2})$
but also $M(R_{1}+R_{2})< M(R_{1})+M(R_{2})$. The configuration with 
maximal entropy and minimal mass is now given by the single black hole.

\section{ Mass gap}
It is evident that for what concerns the entropy of composite 
solutions 2D dilaton gravity models satisfying the conditions (\ref{conda}) and 
(\ref{condb}) of the previous section are very similar to free field theories.
The mass of the gravitational solutions are naturally identified 
with the energy of the excitation in the field theory.
The same is true for generic $d$-dimensional AdS 
solutions when the mass of the gravitational solution is much bigger 
then $L_{p}^{d-2}\la^{d-3}$.
We can ask ourselves if the 2D dilaton gravity models satisfying 
condition (\ref{conda}) but not   (\ref{condb}) have also a field theory counterpart.
More in general, one would like to find field theoretical 
counterparts of $d$-dimensional black hole solutions whose spectrum 
exhibits extremal $M\neq 0$ solutions. 
The most natural candidates are field theories 
with mass gaps. This is rather obvious for the 2D dilaton gravity 
models discussed at the end of Sect. IV (for instance the 
model with exponential dilaton potential), which are 
characterized by a ground state of mass $M_{0}$ with zero entropy and 
nonvanishing temperature. The only way to have a nondegenerate 
ground state at finite temperature is  the presence of  a mass gap  in the 
spectrum. The energy gap $M_{0}$ must be  of the order of the temperature.
In fact for the model with exponential potential discussed in Sect. IV we have 
$E_{gap}=M_{0}\propto T(M_{0})\propto \la$.

The relationship between black hole solutions and field theories  with 
mass gaps seems to be more general. Crucial for the existence of such 
correspondence are both the asymptotic behavior of the metric and the 
presence of an extremal solution with $M_{0}\neq 0$. 
To illustrate this correspondence let us consider  a  generic field 
theory whose spectrum has a mass gap of energy $E_{0}$ separating the 
$E=0$ state from the continuos part of the spectrum $E\ge E_{0}$.
We will also assume that for $E>>E_{0}$ the spectrum is that of a 
generic free field theory and that the entropy of the $E=E_{0}$ state 
vanishes. For $E>>E_{0}$ the entropy/energy relation  will be given 
by Eq. (\ref{enen}) which satisfies  $S(E_{1}+E_{2})<S(E_{1})+S(E_{2})$.
On the other hand for  $E\approx E_{0}$ we have $S=S(E-E_{0})$ with 
$S(E_{0}¥)=0$. Thus, the inequality $S(E_{1}+E_{2}-E_{0})>S(E_{1}-E_{0} )+
S(E_{2}-E_{0})$ will be satisfied at least for $E_{1}=E_{2}=E_{0}$. Because 
for $E>>E_{0}$ the opposite inequality holds,  this implies the 
existence of a threshold energy $\tilde E_{0}$  separating the two 
regimes in complete analogy with what happens for black hole solutions.
This behavior is rather intuitive. For small excitations near $E_{0}$ 
the single state of energy $E=E_{1}+E_{2}$ has more degeneracy then the 
 two  states of energies $E_{1}$ and $E_{2}$ because the 
states below the gap do not contribute. For excitations of energy 
$E>>E_{0}$ the contribution of the gap is irrelevant and the entropy 
is dominated by the contribution coming from the continuos part of 
the spectrum.

\section{ Entropy bounds}

The holographic principle in its usual formulation 
 puts  an upper bound to the entropy, hence to the 
amount of information, which can be stored in given {\sl region of 
space},  $S\le A/4G$, where $A$ is the area of the surface enclosing the region. 
At fixed volume the entropy cannot increase by splitting a 
system into parts, i.e $S(V_{1}+V_{2})\ge S(V_{1})+S(V_{2})$.
This inequality is satisfied not only for extensive systems, when the entropy scales 
as the volume of the  system , but also  when 
the entropy scales as the area of the surface enclosing the system.
An holographic bound cannot be violated by fragmentation. Hence the 
holographic principle is compatible with both Eqs. (\ref{ineq}) and 
(\ref{ineq1}).
This is true for the holographic bound but it is no necessarily true 
for entropy bounds of different type, which can make reference not only to 
the volume  the system (or to the area enclosing it) but also to its energy.
For instance the Bekenstein bound $S\le 2\pi ER$ \cite{Bekenstein:jp} refers not only to 
the linear size $R$ but also to the energy $E$ of the system.

In principle one could also try to formulate entropy bounds, which 
make reference to nothing but the energy of system. 
This could be done following the same line of reasoning one uses to 
formulate the Bekenstein-Hawking bound. In this latter case one  takes 
a system in a {\sl given region} of  space and finds that  
its  entropy is 
bounded by the area of the surface enclosing the region. More in 
detail, when one pumps energy in the system keeping its spatial 
extension  constant,   the entropy increases 
until the energy equals the mass pertaining to a black hole fitting in 
the region. A black hole forms and the bound   $S\le A/4G$ is saturated.

Instead, one could take a system with a  constant {\sl amount of energy} $M$ and look for a 
bound on the associated entropy. Spreading the system over regions of 
space with ever decreasing volumes the entropy will increase until the 
corresponding Schwarzschild radius is reached.   A black hole forms 
and a entropy bound $ S\le \pi GM^{2}$ is found.
The problem is that this bound cannot be universal. Although, the 
Schwarzschild black hole cannot increase its entropy by 
fragmentating, we have seen that there exist  many black hole 
solution that can increase their entropy trough fragmentation.
For this simple reason an entropy bound which refers only to the 
energy of the system cannot be universal.

\section {conclusions}
In this paper we have discussed the holographic principle and entropy 
bounds for composite gravitational systems. We have shown that with 
respect to fragmentation black hole solutions exhibit three different 
types of behavior. The entropy may decrease,  increase or have a mixed 
behavior. In the latter case there will exist a 
threshold mass separating the regime where fragmentation is 
entropically preferred from the regime where it is not. 
Moreover, for 2D black holes we have been  able to find a complete 
characterization of the entropy behavior, in the form of sufficient 
conditions imposed on the function $J$, which defines the 2D 
gravitational model.

Our results have a strong impact on the way one can realize the 
holographic principle. In fact only for gravity theories satisfying 
relation (\ref{ineq}) one can hope to find a realization of the 
holographic principle in terms of a correspondence gravity/field 
theory. For gravity theories satisfying (\ref{ineq1}) this 
correspondence  cannot be realized as a gravity /field theory 
duality and must necessarily have an alternative, still unknown, form.
Because  asymptotically flat black holes, and in particular the 
Schwarzschild black hole, belongs to this latter case we can exclude 
the existence of a correspondence between asymptotically flat black 
holes and a field theory. 

On the other hand we have seen that the class of gravity theories 
satisfying Eq. (\ref{ineq}) is, at least in the 2D case, rather broad.
It contains not only AdS
gravity, for which the gravity/ field theory duality is well 
established, but also other models, for instance those  with 
a exponential dilaton potential. The possibility of finding 
a gravity/ field theory duality for these theories is an open 
question, which deserves further investigations.

An other point of interest is the existence of  black holes with mixed 
behavior, i.e  exhibiting a transition from a regime where 
Eq. (\ref{ineq}) holds to a regime where instead Eq. (\ref{ineq1}) is 
satisfied. We have argued that the most natural candidates for  
holographic duals of this black holes are field theories with a mass 
gap.
Finally, our results seem to exclude the possibility of formulating 
an entropy bound only in terms of the energy of a system: the 
existence of black holes satisfying Eq. (\ref{ineq}) will always 
allow to violate the bound by black hole fragmentation.
 
\begin{acknowledgments}
We thank S. Mignemi for discussions and valuable comments.
\end{acknowledgments}

\end{document}